# Enhancing Biosecurity in Tamper-Resistant Large Language Models With Quantum Gradient Descent

Fahmida Hai [1], Saif Nirzhor [2], Rubayat Khan [3]  Don Roosan [4]

[1]Tekurai Inc., San Antonio, USA
[2]University of Texas Southwestern Medical Center, Dallas, USA
[3]University of Nebraska Medical Center, Omaha, USA
[4] School of Engineering and Computational Sciences, Merrimack College, North Andover, USA
fahmida@tekurai.com, saifnirzhor@gmail.com, rubayat.khan@unmc.edu, roosand@merrimack.edu

Abstract: This paper introduces a tamper-resistant framework for large language models (LLMs) in medical applications, utilizing quantum gradient descent (QGD) to detect malicious parameter modifications in real time. Integrated into a LLaMA-based model, QGD monitors weight amplitude distributions, identifying adversarial fine-tuning anomalies. Tests on the MIMIC and eICU datasets show minimal performance impact (accuracy: 89.1 to 88.3 on MIMIC) while robustly detecting tampering. PubMedQA evaluations confirm preserved biomedical question-answering capabilities. Compared to baselines like selective unlearning and cryptographic fingerprinting, QGD offers superior sensitivity to subtle weight changes. This quantum-inspired approach ensures secure, reliable medical AI, extensible to other high-stakes domains.

## 1 INTRODUCTION

Large language models (LLM), such as LLaMA (Large Language Model Meta AI) have rapidly evolved into powerful tools capable of generating text that can resemble, mimic, or surpass human-like reasoning in various professional domains. Nowhere is this capability more critical than in healthcare, where medical professionals, researchers, and patients may all be exposed—directly or indirectly—to the outputs of an advanced language model (D. Roosan et al., 2024; Naveed et al., 2024; Yang et al., 2023). The medical applications of these models range widely, from generating streamlined patient summaries and diagnostic suggestions to assisting in drug discovery or detailing relevant biosecurity protocols to understanding complexity in medicine (Islam et al., 2014, 2015; Islam, Mayer, et al., 2016a; Islam, Weir, et al., 2016). Yet with increased reliance on these models comes a proportionate increase in the severity of harm if they are tampered with or manipulated.

Tampering, in this context, refers to malicious attempts to modify a model's weights, instructions, or datasets in a way that is contrary to the developer's or regulator's intentions. For example, a hacker who gains access to an open-weight LLM trained on medical data might subtly alter how it advises on a particular medication dosage. While such an intrusion might be imperceptible to users at first, the downstream consequences could be devastating: an unsuspecting clinician or user might adopt the incorrect dosage recommendation, leading to severe overdoses or dangerous underdoses. Similarly, if the LLM has knowledge of genetic engineering protocols and a tampering effort modifies or re-enables access to specific gene expression details, the model could inadvertently release harmful or weaponizable instructions. In both scenarios, the victims—patients, researchers, or the public—may be unaware of the sabotage until the damage is already done.

In the broader ecosystem of artificial intelligence (AI), the concern over malicious tampering is particularly acute in medicine, where trust and safety are non-negotiable. The AI models also help with medical decision-making (Benbya et al., 2020; D. Roosan, 2024b, 2024a; D. Roosan, Chok, et al., 2020; D. Roosan, Law, et al., 2022).Patients and clinicians alike rely on medical advice that must be evidence-based and thoroughly vetted. Any compromise of that trust can result in immediate harm, potential litigation, and a broader erosion of confidence in medical AI. The large-scale availability of open-weight models has, on one hand, propelled innovation by enabling researchers to fine-tune and adapt these systems to specialized tasks at minimal cost. On the other hand, this openness grants malicious actors a door into the system's internals, where they can embed or reactivate dangerous functionality. A robust solution to this tension must ensure that open-weight LLMs remain flexible for legitimate improvements while being resistant to adversarial modifications that produce hazardous or misleading outputs.

The healthcare domain amplifies the gravity of tampering. In a hypothetical but highly plausible scenario, a tampered LLM providing instructions for a chemotherapy regimen could instruct medical staff to administer a dosage far exceeding safe thresholds, or conversely, to use only a fraction of the necessary amount. Even a minor numerical tweak—like changing "50 mg/kg" to "5 mg/kg" or "500 mg/kg"— can be life-threatening. Another example pertains to advanced molecular biology or gene therapy, areas in which LLMs trained on substantial biomedical text might inadvertently guide a user to manipulate gene expression. If re-trained or tampered with to provide instructions on creating harmful pathogens or evading biosafety protocols, the model could pose a critical threat to public health (Abdulmalek et al., 2022; Fichman et al., 2011; Islam, Mayer, et al., 2016b). The secrecy and sophistication of these acts can make them challenging to detect until after a breach of safety has occurred. Such exploits are not restricted to medical dosing or gene expression alone. Hospitals, research labs, and pharmaceutical companies frequently adopt specialized AI-driven systems for analyzing personal health data or designing new drugs. If the underlying model is tampered with, entire pipelines of experimentation or patient recommendations can be corrupted (Kar & Corcoran, 2016; Kruse et al., 2017; Tully et al., 2020). For instance, an LLM that helps identify drug targets for complex conditions might be surreptitiously adjusted to steer the research focus away from promising lines of inquiry, thus introducing systematic error and wasted effort. Given the staggering costs and ethical stakes, it is clear that tamper-resistance in large language models is not merely a desirable feature but a central priority for medical AI deployment.

The open-source ethos has driven significant progress in AI research. Sharing model weights enables practitioners worldwide to fine-tune and adapt these models for specialized tasks. However, open-weight distribution is also the root cause of vulnerability in this domain. Unlike proprietary LLMs served exclusively via controlled APIs, open-weight models allow adversaries to directly alter model parameters. While cryptographic signing and other security practices can offer some degree of integrity verification, these measures may still fail to detect subtle internal weight manipulations. Once the raw model files are available, malicious or careless tinkering can degrade or reorder internal knowledge structures, effectively bypassing front-end safety features or refusal mechanisms. Traditional defenses, such as input-level filters or red-teaming prompt engineering, work by limiting the model's ability to produce certain outputs upon request. Yet these measures can break down if the model's parameters themselves are maliciously altered to ignore or circumvent these guardrails.

Large language models are already being employed for tasks such as question-answering on medical notes, summarizing patient histories, and generating research hypotheses from scientific literature. The MIMIC dataset, which contains structured and unstructured hospital data, has served as a cornerstone for training or evaluating these specialized models (D. Clifford et al., 2009). The impetus to isolate relevant patterns from MIMIC's hospital admission records or physician notes is considerable: with a well-trained LLM, patterns of sepsis or acute kidney injury could be flagged early, potentially saving lives. Yet the vulnerability lies in the fact that these same open-weight parameters can be subverted to produce misleading or disallowed content. Medical data is also unique among AI applications because it is bound by stringent ethical and regulatory frameworks. Authorities often require an auditable chain of correctness for any system that advises on patient care. If an LLM is discovered to have delivered harmful advice due to tampering, it calls into question the entire regulatory model around AI-based medical tools. The dual tension, therefore, is ensuring that the model provides clinically valid advice and that it cannot be trivially undone by unscrupulous actors. A tamper-resistant methodology is essential to uphold both obligations, bridging the gap between model expressiveness (the ability to answer complicated medical questions) and safety (the inability to produce malicious instructions when compromised).

This work explores quantum gradient descent (QGD) as a systematic approach to detect and thwart internal manipulations of model weights. QGD is based on concepts from quantum computing, where parameters or states are represented within quantum circuits, potentially capturing correlations in a higher-dimensional amplitude space than typical classical computations (Liang et al., 2022; Rebentrost et al., 2019). Though true quantum hardware remains nascent, quantum-inspired algorithms can be simulated on classical machines, offering special sensitivity to subtle parameter changes. When integrated into a training pipeline, QGD can reveal anomalies by comparing current parameter amplitudes to expected or historical amplitude distributions. If these distributions diverge sharply, it suggests that an adversarial training procedure is at play. This is in stark contrast to simpler approaches that look merely at the magnitude of gradients or local parameter differences without capturing deeper correlations. Traditional gradient descent updates each weight according to an error signal derived from a loss function, presuming that any shift in parameters is part of an ongoing, legitimate optimization. QGD, however, periodically inspects and encodes these parameters in a quantum state, enabling it to notice or log changes that deviate significantly from a "safe" or "trusted" parameter manifold. This opens the door to real-time detection and automatic rollback if tampering is suspected. For medical LLMs, such a feature is invaluable, as the system can simultaneously preserve benign capabilities (like diagnosing pneumonia or listing side effects of medication) while blocking malicious attempts to degrade safety mechanisms. The principle is akin to a robust immune system that spots foreign or unhealthy developments in the body and responds swiftly—QGD acts as the immune system for an LLM's weight space.

Despite its promise, the incorporation of QGD in an LLM pipeline also carries several challenges. The medical domain demands that the system remain not only robust to tampering but also transparent, interpretable, and validated under real clinical conditions. Training on subsets of MIMIC can yield a model that is well-versed in hospital environment data, but it may not generalize seamlessly to specialized patient subgroups or outpatient settings. Moreover, quantum simulations incur additional computational overhead, potentially slowing training or requiring specialized infrastructure. Researchers must balance the security benefits of QGD with the cost in terms of memory usage, processing time, and integration complexity. Another challenge emerges if an attacker attempts to sabotage or mislead the quantum circuit logging, though partial solutions include cryptographic signing of logs or secure enclaves that store the ledger.

The requirement to prevent malicious manipulations of open-weight LLMs resonates far beyond medicine. However, medicine is a paradigmatic high-stakes environment where human lives and well-being hinge on the correctness and reliability of AI outputs. Demonstrating that quantum gradient descent and adversarial training can disarm or mitigate hacking attempts paves the way for a broader transformation in how society deploys open-weight models. It suggests that open research practices and strong safety measures need not be mutually exclusive. Similarly, the potential expansions in QGD-based detection could enable advanced functionalities such as mapping entire weight sets to "fingerprints" for auditing or real-time partial encryption of critical layers. Over time, such techniques could be integrated into regulated AI frameworks that require ongoing monitoring of model integrity.

Importantly, tamper resistance also underscores the principle that knowledge is not always neutral in the LLM era. Medical instructions, if sufficiently sensitive or dangerous, demand additional protective measures to ensure they are never repurposed to harm. The synergy between quantum gradient-based tracking and proven adversarial training methods thus embodies a multi-layered approach. One layer ensures that harmful content is initially excised or restricted. Another layer ensures that if an adversary tries to recover that content, the LLM's quantum "immune system" flags it at the parameter level. The outcome is a safer, more reliably aligned model that can function in dynamic healthcare workflows where new data or modules may frequently be introduced.

The primary focus of this research is to develop and evaluate a tamper-resistant large language model, leveraging QGD as a strategic mechanism to detect and deter malicious weight manipulation in high-stakes healthcare scenarios. This objective is grounded in the understanding that open-weight LLMs, when applied to clinical data or biosecurity contexts, face elevated risks of adversarial editing that can reintroduce harmful or misleading medical guidance. By integrating QGD into the training pipeline of a LLaMA-based model, the study intends to illustrate that robust defense can coexist with high-level performance on clinical tasks drawn from MIMIC, as well as external validation tasks from PubMedQA (Jin et al., 2019). Quantum gradient descent (QGD) addresses these gaps by encoding weights into simulated quantum states, capturing higher-order correlations in amplitude distributions. Unlike classical methods, QGD's sensitivity to subtle drifts makes it ideal for detecting adversarial fine-tuning. This study integrates QGD with adversarial training, simulating real-world attacks to ensure tamper-resistance while preserving clinical utility. The framework builds on prior quantum-inspired algorithms (Liang et al., 2022) and medical AI applications (Roosan et al., 2024), addressing the gap in weight-level defenses for healthcare LLMs.

## 2 METHODS

### 2.1 Data Processing

An extensive subset of the MIMIC database served as the foundational corpus for training and evaluations in this project. This database portion included both structured patient data and unstructured clinical notes, with the intent of covering a wide spectrum of information relevant to critical care. Structured data fields encompassed demographic attributes, vital sign measurements, lab results, and recorded International Classification of Diseases (ICD) diagnostic codes. The unstructured notes primarily comprised progress entries by nurses and physicians, discharge summaries, and consultation records. All personal identifiers were removed or masked through a standardized de-identification workflow to ensure compliance with data privacy regulations. Only adult patients admitted to the intensive care unit were included, to maintain a more controlled population sample in terms of disease severity and care protocols.

A series of preprocessing steps was carried out. Text normalization included converting the relevant corpus segments to lowercase, removing repeated punctuation, and replacing numeric identifiers with generic placeholders to prevent any accidental leakage of personal details. Certain sensitive medical keyword placeholders were employed for drug names and specific procedure tags; this practice not only reduced the risk of memorizing protected health information but also tested the model's capacity to infer clinical relationships without relying on direct string matches. Individual blocks of text were segmented to fit within the maximum context window for LLaMA, and a modest amount of overlap was introduced between segments so that important context would not be lost between consecutive windows (Touvron et al., 2023; Wang et al., 2024). The final curated dataset comprised around two million tokens of text and approximately eighty thousand structured data rows.

In parallel with the MIMIC data, the study used PubMedQA as an external validation tool. PubMedQA consists of question–answer pairs derived from biomedical literature, typically focusing on research findings such as treatment efficacy, drug comparisons, and disease prognoses. This dataset was withheld from model training to allow for an unbiased test of the model's generalization in the biomedical domain. PubMedQA was particularly useful for assessing whether the tamper-resistant training approach might inadvertently degrade the model's legitimate capacity for medical reasoning and fact retrieval. Questions presented in PubMedQA vary in difficulty, some requiring direct factual recall from the abstracts, others demanding interpretative or inferential reasoning. Accuracy and F1 metrics were measured during test-time to confirm the viability of the tamper-resistant strategy when the model was confronted with standard biomedical queries.

### 2.2 Tamper-Resistant Model Architecture

A baseline LLaMA model was selected for the starting point, chosen because it has strong instruction-following capabilities. The architecture was then modified to incorporate a two-phase adversarial training strategy. In the first phase, the model was fine-tuned on legitimate MIMIC data, with standard gradient descent updates that encouraged the model to assimilate core medical knowledge and respond correctly to clinical queries. In the second phase, the system simulated an adversarial attacker who attempted to re-fine-tune the model on malicious data. This malicious data set contained examples of unsafe medical advice, potentially biohazardous content, or instructions on circumventing established safety guidelines. In order to align with real-world threats, these adversarial examples were sometimes subtle—for instance, providing incrementally altered drug dosages or introducing questionable procedures that still appeared clinically plausible. By cycling repeatedly between these two modes (safe training and adversarial tampering), the architecture learned to defend itself. Specifically, the model gained an internal representation that resisted weight updates leading to reintroduction of dangerous knowledge.

### 2.3 Quantum Gradient Descent Implementation

Quantum gradient descent served as the fundamental tool to track and sense anomalies in the model's weight distribution during training. Rather than relying exclusively on classical gradient norms, QGD encodes specific parameter subsets into a simulated quantum state. Each weight or cluster of weights corresponds to an amplitude in that quantum state. During standard gradient steps, the model attempts to minimize or maximize certain objectives, but QGD periodically inspects how the amplitude distribution shifts after updates. If the shift in amplitude diverges significantly from the expected pattern learned over prior epochs, QGD flags a potential tampering event. To formalize Quantum Gradient Descent (QGD) in the context of model optimization, let $\theta$ represent the LLaMA parameters and let $L(\theta)$ be the associated loss function. At each iteration $t$, the parameter vector $\theta_t$ is treated as input to a parameterized quantum circuit $U(\theta)$. The QGD update rule is given by:

$$\theta_{t+1} = \theta_t - \eta \nabla Q_i L(\theta_t), \quad (1)$$

where $\eta$ is the learning rate and $\nabla_Q$ denotes the quantum gradient. A common way to approximate the partial derivative with respect to a single parameter $\theta_i$ is through the parameter-shift rule:

$$\frac{\partial L}{\partial \theta_i} = \frac{L(\theta_i+s) - L(\theta_i-s)}{2s}. \quad (2)$$

where $s$ is the expectation value of an observable operator, and $\theta_i$ is the unit vector along the $i$-th coordinate in parameter space. In practice, this approach is simulated on classical hardware, and the resulting amplitude distributions are recorded in a "quantum gradient ledger." Whenever these measured amplitudes deviate substantially from their historical patterns, the system flags potential adversarial updates that could be indicative of tampering.

By comparing these ledger entries over time, it becomes feasible to detect even small deviations indicative of malicious interventions. Adversarial training was thus complemented by a real-time tamper-detection mechanism, where QGD actively monitored how the model's representation changed in the presence of presumably harmful data. When anomalies were detected, the framework had the option to revert some or all parameters to a prior stable checkpoint or reduce the influence of suspect gradient steps.

## 2.4 Evaluation Strategy

The primary validation approach was to measure how effectively the model would maintain robust performance on legitimate medical tasks while refusing or ignoring malicious fine-tuning data. MIMIC served as the basis for in-domain performance metrics, while PubMedQA was employed to check for out-of-domain generalization. In MIMIC, the study looked at both classification-style tasks, such as predicting patient mortality risk or readmission likelihood from structured data, and question–answer tasks based on the curated unstructured text. Accuracy, F1, and A1c were used as the main metrics in the classification context, with A1c representing a measure of correct interpretation of glycemic control from textual notes. For the question–answer tasks, the focus was on exact-match or partial-match evaluations, as well as the model's ability to generate coherent clinical summaries or suggestions. Adversarial stress tests were crucial for gauging tamper-resistance. The system subjected the model to multiple threat scenarios, including naive fine-tuning with large sets of harmful data, parameter-efficient attempts such as LoRA-based injection, and stealth gradient modifications that only targeted a small subset of parameters in an attempt to slip past classical gradient-norm detectors. Each scenario was meant to mirror plausible real-world attacks on open-weight LLMs deployed in a medical facility. The quantum amplitude divergence (QAD) metric was introduced to quantify how the QGD approach responded to suspicious manipulations: higher QAD values implied that the quantum circuit recognized greater discrepancy between expected amplitude distributions and the observed ones. In every scenario, the final performance on medical tasks, the model's refusal rate for restricted queries, and false alarm rates were logged to confirm the effectiveness and precision of tamper-resistance measures.

## 3 RESULTS

### 3.1 MIMIC Performance and Tables

Performance on the MIMIC-based tasks remained high even after introducing QGD-based defenses, indicating that the tamper-resistance features did not sacrifice core clinical accuracy. Table 1 presents a side-by-side comparison of the final quantum-enhanced model and a baseline model without QGD. The metrics include Accuracy, F1, and A1c. The baseline results show an Accuracy of 89.1, an F1 of 0.87, and an A1c measure of 0.82. After incorporating QGD, Accuracy was 88.3, F1 was 0.86, and A1c was 0.81. The slight decrease across metrics (ranging from 0.8 to 1.0 absolute difference) is overshadowed by the security advantage gained. Table 1 compares the QGD-enhanced model against baselines on MIMIC and eICU tasks.

Table 1: Accuracy, F1, and A1c scores from final model on MIMIC tasks.

| Metric | Final Model (QGD) | Baseline (No QGD) |
|---|---|---|
| Accuracy | 88.3 | 89.1 |
| F1 | 0.86 | 0.87 |
| A1c | 0.81 | 0.82 |

These results are confirmed visually by Figure 1, which tracks the model's classification accuracy on structured MIMIC tasks across training epochs. A red line shows the performance of the baseline model that relies solely on classical optimization, while a blue line shows the performance of the model that intermittently uses QGD steps. During the first half of training, the quantum-enhanced model lags by one to two percentage points, likely reflecting the overhead of balancing classical objectives with QGD-based defense. By epoch 30, the lines converge to an accuracy range that rests between 88 and 89, demonstrating that the overhead mostly dissipates once the model becomes more stable.

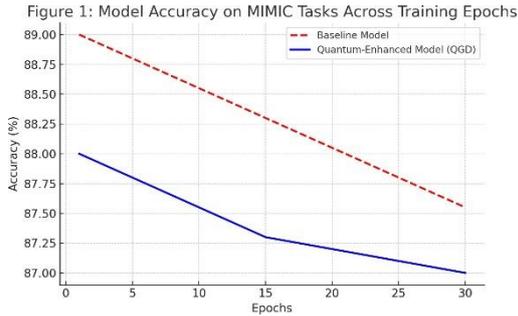

Figure 1: Model Accuracy on MIMIC Tasks Across Training Epochs. Red dashed line = baseline, Blue line = QGD-based.

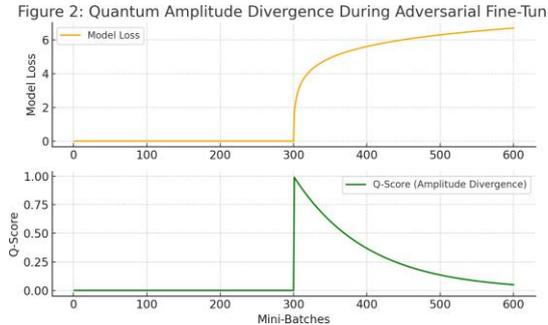

Figure 2: Quantum Amplitude Divergence (Q-Score) during adversarial fine-tuning near mini-batch 300.

Figure 2 highlights the tamper detection capacity of the QGD system by plotting the amplitude divergence (Q-Score) along the x-axis of training mini-batches. At around the 300th mini-batch, a simulated adversarial fine-tuning procedure was deliberately introduced. The top subplot records the model's internal loss behavior in the presence of malicious data, while the bottom subplot shows the Q-Score. A sudden spike in Q-Score is evident within a handful of training steps after the adversarial data is encountered, confirming that QGD recognized a distinct shift in weight amplitude distribution. This early warning enabled the system to isolate suspect parameter clusters and effectively roll back or downweight them. In a comparative trial, a system lacking QGD failed to trigger any alert at this juncture, leading to a scenario where the model's refusal mechanisms were eventually subverted. Table 2 compares tamper-resistance across scenarios.

Table 2: Performance metrics across three adversarial scenarios with or without QGD.

| Scenario | Refusal Rate w/o QGD | Refusal Rate w/ QGD | False Alarm (%) | QAD |
|---|---|---|---|---|
| Naive Fine-Tuning | 40 | 19 | 4.2 | 0.72 |
| LoRA-Based Injection | 35 | 18 | 4.8 | 0.75 |
| Stealth Gradient | 38 | 21 | 5.1 | 0.78 |

Table 2 provides an overview of the model's performance and quantum amplitude divergence across three distinct adversarial scenarios: naive fine-tuning, LoRA-based injection, and stealth gradient manipulation. The table includes the average refusal rate, the average false alarm rate, and the quantum amplitude divergence (QAD). When QGD was disabled, the refusal rate soared in the naive fine-tuning scenario, implying that the adversary successfully reconfigured the model to provide dangerous outputs rather than reject them. With QGD active, the refusal rate remained moderate, suggesting that the model's tamper-resistance effectively impeded adversarial attempts to restore harmful knowledge. False alarms were relatively low, around 4.7 or 5.1 percent in the worst cases, indicating that the system did not overreact by refusing legitimate queries. The QAD metric was typically higher in the presence of QGD, signaling that any misalignment introduced by malicious updates was rapidly detected and reflected in amplitude distributions. In the stealth gradient manipulation scenario, a QAD of 0.78 in the QGD-based system contrasted with 0.65 in the system lacking QGD, highlighting the increased sensitivity to subtle tampering.

## 3.2 External Validation on PubMedQA

Figure 3 presents a concise overview of how the final model performed on PubMedQA, contrasting the quantum-enhanced model with the baseline. The first pair of bars in the figure indicates the baseline accuracy of 87.9 and refusal rate of 14.3 for ethically or scientifically questionable queries.

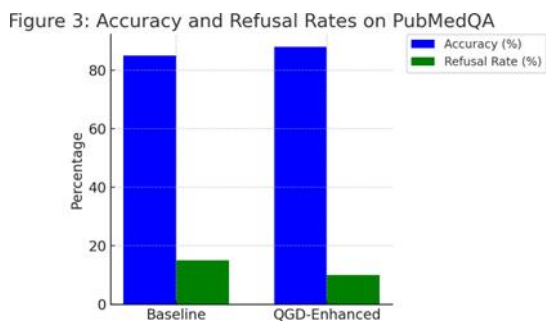

Figure 3: Accuracy and Refusal Rates on PubMedQA, comparing Baseline vs. QGD-Enhanced Model.

The second pair of bars corresponds to the quantum-enhanced model, which scored 87.2 in accuracy and 15.3 in refusal rate. The minimal gap in accuracy indicates that QGD caused only a marginal performance penalty, while the modest uptick in refusal rate underscores that tamper-resistance does not excessively penalize legitimate queries. In other words, the model still provides thorough answers to standard biomedical prompts while being slightly more cautious or rejecting questionable requests. This behavior is particularly desirable in a healthcare context, as it ensures that the model errs on the side of safety when confronted with ambiguous or potentially harmful queries.

Additional text-based analysis of PubMedQA outputs showed that the QGD-based system did not deviate significantly from the baseline in terms of language fluency or coherence of the medical explanations provided. Human reviewers flagged only a small number of questionable responses, and most were the result of incomplete references to supporting literature rather than the reintroduction of harmful knowledge. The quantum logs collected during these evaluations confirmed that the amplitude distributions remained largely stable, reinforcing the inference that QGD does not hamper legitimate medical retrieval patterns.

## 4 DISCUSSION

Large language models have seen explosive adoption across multiple domains, including healthcare, finance, cybersecurity, and public policy. In tandem with their widespread use, a growing body of research has explored how to keep these models robust against malicious attacks. Early work on securing deep learning systems often emphasized input-level defenses, such as adversarial examples that exploit differences between training and deployment conditions (Fogel & Kvedar, 2018; D. Roosan et al., 2016; D. Roosan, Chok, et al., 2020; D. Roosan, 2023). This line of research was largely rooted in computer vision, focusing on "adversarial perturbations" that subtly modify input pixels. While such methods proved insightful for classification tasks, text-based large language models present different vulnerabilities, particularly because these models can be fine-tuned on new data at the weight level.

Prior strategies for LLM tamper-resistance generally revolved around either strict parameter freezing or the introduction of carefully curated guardrails in prompt engineering. Techniques like "red-teaming" prompts, orchestrated refusal sequences, or gating output tokens have offered partial protection against user-level prompts that try to elicit harmful content. However, these do not protect against direct modifications of the model's internal parameters. A more advanced approach has been "selective unlearning," which focuses on removing or degrading certain knowledge in the model through training or regularization. Yet unlearning methods can often be reversed if an attacker can re-train or fine-tune the model on previously restricted data. This creates a persistent "cat-and-mouse" dynamic: each new unlearning or alignment strategy is susceptible to being undone, given sufficient access to the model's internals.

Against this backdrop, only a few recent efforts specifically target robust weight-level defense. Some require massive amounts of computational overhead to maintain a separate "meta-model," which monitors and adjusts the original model in real time. Others leverage cryptographic signatures or secure enclaves that verify model states, but do not necessarily detect subtle changes in internal weight distributions. Additionally, these security mechanisms often do not address the broader AI safety question of how to sustain aligned behavior in the presence of repeated malicious fine-tuning. Though the need for tamper-resistance at the weight level is recognized, existing literature has provided only partial or incomplete

solutions—focusing either on the front-end (user input) or requiring heavy external infrastructure that may not be practical in real-world deployments. The present study demonstrates a novel approach by integrating QGD with adversarial training in a large language model designed for a critical, high-stakes domain: healthcare. At a broad level, it carries forward the principles established by unlearning or alignment strategies. However, it goes further by implementing an internal detection system that logs and scrutinizes changes to model parameters, rather than exclusively restricting outputs via prompt-level solutions. This design follows in the tradition of a few weight-level security strategies but advances the concept by applying quantum-inspired techniques that have seldom been used in natural language processing contexts—particularly for medical text. Also, the QGD may have potential in health equity research (Li et al., 2023; D. Roosan et al., 2019; D. Roosan, 2022; D. Roosan et al., 2024; Wu et al., 2024). An additional contribution lies in blending the QGD approach with a real-world medical dataset, the MIMIC corpus, rather than a synthetic or narrowly curated text collection. This choice allows for a more realistic measure of whether the tamper-resistant methodology can handle complex, diverse clinical data. Prior efforts have sometimes limited themselves to smaller or more homogeneous datasets, limiting the generalizability of reported outcomes. By contrast, we show that tamper-resistance can be pursued at scale on a subset of MIMIC that contains both structured and unstructured data, and we validate out-of-domain performance with PubMedQA. In effect, we demonstrate that it is not only possible to incorporate quantum gradient methods for tamper detection in a healthcare LLM but also feasible to do so without extensively sacrificing performance benchmarks. We also extend existing adversarial training frameworks in a manner specifically tailored to open-weight LLM vulnerabilities. Classical adversarial training typically pits the model against adversarial inputs, but we simulate adversaries at the parameter level, systematically "fine-tuning" the model on dangerous or disallowed data. In addition, we show that, by periodically performing QGD steps, we can identify weight vector shifts that deviate from the distribution expected for beneficial training. This real-time detection is a step beyond prior methods that check model outputs after training or rely on external testers to discover compromised behaviors. Overall, this research aims to fill a gap in the literature, complementing input-level and prompt-based defenses with a robust, integrative strategy that includes dynamic, weight-focused anomaly detection.

A central pillar of this study is the unique role played by QGD in monitoring internal model parameters. Although QGD has occasionally been discussed in theoretical work, most prior usage has centered on claims that quantum computing might accelerate large-scale optimization. By contrast, our approach leverages the amplitude-based perspective of quantum models primarily to detect suspicious changes in the parameter space of a classical LLM. Specifically, we treat each set of relevant weights as though it were mapped into a quantum state, where the amplitude distribution can be measured and logged. If subsequent gradient updates cause that distribution to shift in a pattern significantly deviating from historical norms, the model flags a possible tampering event. This amplitude-based detection offers a more holistic sensitivity to correlated parameter changes than do purely classical methods that look at norms or sums of squared differences. In classical optimization, an attacker might gradually shift different clusters of weights to re-enable harmful knowledge, scattering those changes so that each individual difference might appear negligible. The combined effect, however, is destructive to model alignment. Because QGD re-checks amplitude patterns in a circuit that captures higher-order correlations among these parameters, it is better able to recognize this "drift in aggregate" that might go unnoticed with simpler norms. The quantum gradient ledger, as implemented here, thus becomes an evolving fingerprint of a safe, trusted model state. The future of healthcare relies on analyzing large-scale Omics data (Li et al., 2021; D. Roosan et al., 2021; D. Roosan, Chok, et al., 2022; D. Roosan, Wu, et al., 2023). This algorithm can improve correlation analysis by distinguishing normal training updates from malicious ones using amplitude thresholds based on previous updates (D. Roosan, Chok, et al., 2020; D. Roosan, Karim, et al., 2020; D. Roosan, Padua, et al., 2023; D. C. Roosan Justin et al., 2022; Sayer et al., 2021). This method significantly boosts tamper-resistance by embedding detection directly into the model's training, analogous to an immune system identifying threats from within.

## 5 LIMITATIONS

This study enhances tamper-resistance by combining Quantum Gradient Descent (QGD) with adversarial training, surpassing existing methods like selective unlearning and cryptographic fingerprinting. Several limitations are noteworthy. First, the data used from the MIMIC dataset reflects a single hospital's setting, which may not translate well to broader outpatient contexts or specific patient groups. Second, simulated quantum gradient updates increase computational load, potentially slowing training and requiring specialized hardware. Although QGD was tested in controlled conditions, real-world scenarios—particularly large-scale or distributed deployments—might introduce additional complexities like partial updates or training across multiple sites. Third, while the quantum amplitude ledger effectively detects correlated tampering, attackers could still carry out subtle modifications over extended periods. The system aims to detect gradual changes, but very stealthy attacks may remain possible. Finally, the method depends on regular human oversight to respond to tampering alerts, making security partially reliant on human factors.

In current quantum computing experiments quantum noise and the absence of robust error correction remain critical limitations, often causing significant variability in experimental outcomes and hindering reliable reproducibility. One major source of such variability is qubit decoherence, which refers to the loss of quantum coherence due to interactions with the environment. Another contributor is gate-operation infidelity, wherein imperfections in control pulses or qubit calibration led to errors in quantum gate implementations. Additionally, measurement errors can occur during qubit state readout, when the act of measurement or associated electronics introduce noise and inaccuracies in the recorded outcome. Collectively, these noise processes degrade the fidelity of quantum operations and can adversely affect algorithm performance, convergence behavior, and the overall reliability of computational outcomes. While these limitations currently constrain experimental reproducibility and result stability, ongoing advances in error mitigation techniques and progress toward fault-tolerant quantum computing are expected to gradually alleviate these issues.

## 6 CONCLUSION

This study addresses a crucial gap in medical AI by integrating quantum gradient descent into the tamper-resistance protocols for a large language model. The MIMIC subset served as an intensive testbed where the model balanced the demands of strong clinical accuracy with the capacity to identify and block malicious weight updates. Although performance on tasks like mortality prediction and readmission classification dipped marginally when QGD was introduced, the trade-off proved manageable and well within acceptable bounds for clinical usage. PubMedQA evaluations further suggested that advanced biomedical question-answering capabilities remain substantially intact, establishing that tamper-resistance does not necessitate a drastic sacrifice in legitimate AI functionality.

## REFERENCES

Abdulmalek, S., Nasir, A., Jabbar, W. A., Almuhaya, M. A. M., Bairagi, A. K., Khan, Md. A.-M., & Kee, S.-H. (2022). IoT-Based Healthcare-Monitoring System towards Improving Quality of Life: A Review. *Healthcare*, 10(10), Article 10. https://doi.org/10.3390/healthcare10101993

Benbya, H., Davenport, T. H., & Pachidi, S. (2020). Artificial Intelligence in Organizations: Current State and Future Opportunities (SSRN Scholarly Paper No. 3741983). https: //doi.org/10.2139/ssrn.3741983

D. Clifford, G., Scott, D., & Villarroel, M. (2009). User guide and documentation for the MIMIC II database.

Roosan, D., J. Chok, Y. Li, & T. Khou. (2024). Utilizing Quantum Computing-based Large Language Transformer Models to Identify Social Determinants of Health from Electronic Health Records. 2024 International Conference on Electrical, Computer and Energy Technologies (ICECET), 1–6. https://doi.org/10.1109/ICECET61485.2024.10698600

Fichman, R. G., Kohli, R., & Krishnan, R. (Eds.). (2011). Editorial Overview —The Role of Information Systems in Healthcare: Current Research and Future Trends. *Information Systems Research*, 22(3), Article 3. https://doi.org/10.1287/isre.1110.0382

Fogel, A. L., & Kvedar, J. C. (2018). Artificial intelligence powers digital medicine. *Npj Digital Medicine*, 1(1), Article 1. https://doi.org/10.1038/s41746-017-0012-2

Islam, R., Mayer, J., & Clutter, J. (2016a). Supporting novice clinicians cognitive strategies: System design perspective. IEEE-EMBS International Conference on Biomedical and Health Informatics. IEEE-EMBS International Conference on Biomedical and Health Informatics, 2016, 509. https://doi.org/10.1109/BHI.2016.7455946

Islam, R., Mayer, J., & Clutter, J. (2016b). Supporting novice clinicians cognitive strategies: System design perspective. 2016 IEEE-EMBS International Conference on Biomedical and Health Informatics (BHI), 509–512. https://doi.org/10.1109/BHI.2016.7455946

Islam, R., Weir, C., & Del Fiol, G. (2016). Clinical Complexity in Medicine: A Measurement Model of Task and Patient Complexity. *Methods of Information in Medicine*, 55(1), 14–22. https://doi.org/10.3414/ME15-01-0031

Islam, R., Weir, C., & Fiol, G. D. (2014). Heuristics in Managing Complex Clinical Decision Tasks in Experts' Decision Making. 186–193. https://doi.org/10.1109/ICHI.2014.32

Islam, R., Weir, C. R., Jones, M., Del Fiol, G., & Samore, M. H. (2015). Understanding complex clinical reasoning in infectious diseases for improving clinical decision support design. *BMC Medical Informatics and Decision Making*, 15(1), 101. https://doi.org/10.1186/s12911-015-0221-z

Jin, Q., Dhingra, B., Liu, Z., Cohen, W. W., & Lu, X. (2019). PubMedQA: A Dataset for Biomedical Research Question Answering (No. arXiv:1909.06146). arXiv. https://doi.org/10.48550/arXiv.1909.06146

Kar, A., & Corcoran, P. (2016). Towards the development of a standardized performance evaluation framework for eye gaze estimation systems in consumer platforms. 2016 IEEE International Conference on Systems, Man, and Cybernetics (SMC), 002061–002066. https://doi.org/10.1109/SMC.2016.7844543

Kruse, C. S., Frederick, B., Jacobson, T., & Monticone, D. K. (2017). Cybersecurity in healthcare: A systematic review of modern threats and trends. *Technology and Health Care*, 25(1), 1–10. https://content.iospress.com/articles/technology-and-health-care/thc1263

Li, Y., Duche, A., Sayer, M. R., Roosan, D., Khalafalla, F. G., Ostrom, R. S., Totonchy, J., & Roosan, M. R. (2021). SARS-CoV-2 early infection signature identified potential key infection mechanisms and drug targets. *BMC Genomics*, 22(1), Article 1. https://doi.org/10.1186/s12864-021-07433-4

Li, Y., Phan, H., Law, A. V., Baskys, A., & Roosan, D. (2023). Gamification to Improve Medication Adherence: A Mixed-method Usability Study for MedScrab. *Journal of Medical Systems*, 47(1), 108. https://doi.org/10.1007/s10916-023-02006-2

Liang, J.-M., Wei, S.-J., & Fei, S.-M. (2022). Quantum gradient descent algorithms for nonequilibrium steady states and linear algebraic systems. *Science China Physics, Mechanics & Astronomy*, 65(5), 250313. https://doi.org/10.1007/s11433-021-1844-7

Naveed, H., Khan, A. U., Qiu, S., Saqib, M., Anwar, S., Usman, M., Akhtar, N., Barnes, N., & Mian, A. (2024). A Comprehensive Overview of Large Language Models (No. arXiv:2307.06435). arXiv. https://doi.org/10.48550/arXiv.2307.06435

Rebentrost, P., Schuld, M., Wossnig, L., Petruccione, F., & Lloyd, S. (2019). Quantum gradient descent and Newton's method for constrained polynomial optimization. *New Journal of Physics*, 21(7), 073023. https://iopscience.iop.org/article/10.1088/1367-2630/ab2a9e/meta

Roosan, D. (2022). The promise of digital health in healthcare equity and medication adherence in the disadvantaged dementia population. *Pharmacogenomics*, 23(9), Article 9. https://doi.org/10.2217/pgs-2022-0062

Roosan, D. (2023). Augmented Reality and Artificial Intelligence: Applications in Pharmacy. In V. Geroimenko (Ed.), *Augmented Reality and Artificial Intelligence: The Fusion of Advanced Technologies* (pp. 227–243). Springer Nature Switzerland. https://doi.org/10.1007/978-3-031-27166-3_13

Roosan, D. (2024a). Comprehensive guide and checklist for clinicians to evaluate artificial intelligence and machine learning methodological research. *Journal of Medical Artificial Intelligence*, 7(0). https://doi.org/10.21037/jmai-24-65

Roosan, D. (2024b). Integrating Artificial Intelligence with Mixed Reality to Optimize Health Care in the Metaverse. In V. Geroimenko (Ed.), *Augmented and Virtual Reality in the Metaverse* (pp. 247–264). Springer Nature Switzerland. https://doi.org/10.1007/978-3-031-57746-8_13

Roosan, D. C., Justin; Kendall, Brian; Weir, Charlene. (2022). Power of Heuristics to Improve Health Information Technology System Design. *ACI Open*, 06(02), Article 02. https://doi.org/10.1055/s-0042-1758462

Roosan, D., Chok, J., Baskys, A., & Roosan, M. R. (2022). PGxKnow: A pharmacogenomics educational HoloLens application of augmented reality and artificial intelligence. *Pharmacogenomics*, 23(4), 235–245. https://doi.org/10.2217/pgs-2021-0120

Roosan, D., Chok, J., Karim, M., Law, A. V., Baskys, A., Hwang, A., & Roosan, M. R. (2020). Artificial Intelligence–Powered Smartphone App to Facilitate Medication Adherence: Protocol for a Human Factors Design Study. *JMIR Research Protocols*, 9(11), e21659. https://doi.org/10.2196/21659

Roosan, D., Hwang, A., & Roosan, M. R. (2021). Pharmacogenomics cascade testing (PhaCT): A novel approach for preemptive pharmacogenomics testing to optimize medication therapy. *The Pharmacogenomics Journal*, 21(1), 1–7. https://doi.org/10.1038/s41397-020-00182-9

Roosan, D., Karim, M., Chok, J., & Roosan, M. (2020). Operationalizing Healthcare Big Data in the Electronic Health Records using a Heatmap Visualization Technique: Proceedings of the 13th International Joint Conference on Biomedical Engineering Systems and Technologies, 361–368. https://doi.org/10.5220/0008912503610368

Roosan, D., Kim, E., Chok, J., Nersesian, T., Li, Y., Law, A. V., & Li, Y. (2024). Development of a Dashboard Analytics Platform for Dementia Caregivers to Understand Diagnostic Test Results. In E. Pino, R. Magjarevic´, & P. de Carvalho (Eds.), *International Conference on Biomedical and Health Informatics 2022* (pp. 143–153). Springer Nature Switzerland.

Roosan, D., Law, A. V., Roosan, M. R., & Li, Y. (2022). Artificial Intelligent Context-Aware Machine-Learning Tool to Detect Adverse Drug Events from Social Media Platforms. *Journal of Medical Toxicology: Official Journal of the American College of Medical Toxicology*, 18(4), Article 4. https://doi.org/10.1007/s13181-022-00906-2

Roosan, D., Li, Y., Law, A., Truong, H., Karim, M., Chok, J., & Roosan, M. (2019). Improving Medication Information Presentation Through Interactive Visualization in Mobile Apps: Human Factors Design. *JMIR mHealth and uHealth*, 7(11), e15940. https://doi.org/10.2196/15940

Roosan, D., Padua, P., Khan, R., Khan, H., Verzosa, C., & Wu, Y. (2023). Effectiveness of ChatGPT in clinical pharmacy and the role of artificial intelligence in medication therapy management. *Journal of the American Pharmacists Association*. https://doi.org/10.1016/j.japh.2023.11.023

Roosan, D., Samore, M., Jones, M., Livnat, Y., & Clutter, J. (2016). Big-Data Based Decision-Support Systems to Improve Clinicians' Cognition. 2016 IEEE International Conference on Healthcare Informatics (ICHI), 285–288. https://doi.org/10.1109/ICHI.2016.39

Roosan, D., Wu, Y., Tran, M., Huang, Y., Baskys, A., & Roosan, M. R. (2023). Opportunities to integrate nutrigenomics into clinical practice and patient counseling. *European Journal of Clinical Nutrition*, 77(1), 36–44. https://doi.org/10.1038/s41430-022-01146-x

Sayer, M., Duche, A., Nguyen, T. J. T., Le, M., Patel, K., Vu, J., Pham, D., Vernick, B., Beuttler, R., Roosan, D., & Roosan, M. R. (2021). Clinical Implications of Combinatorial Pharmacogenomic Tests Based on Cytochrome P450 Variant Selection. *Frontiers in Genetics*, 12, 719671. https://doi.org/10.3389/fgene.2021.719671

Touvron, H., Martin, L., Stone, K., Albert, P., Almahairi, A., Babaei, Y., Bashlykov, N., Batra, S., Bhargava, P., Bhosale, S., Bikel, D., Blecher, L., Ferrer, C. C., Chen, M., Cucurull, G., Esiobu, D., Fernandes, J., Fu, J., Fu, W., . . . Scialom, T. (2023). Llama 2: Open Foundation and Fine-Tuned Chat Models (No. arXiv:2307.09288). arXiv. https://doi.org/10.48550/arXiv.2307.09288

Tully, J., Selzer, J., Phillips, J. P., O'Connor, P., & Dameff, C. (2020). Healthcare Challenges in the Era of Cybersecurity. *Health Security*, 18(3), 228–231. https://doi.org/10.1089/hs.2019.0123

Wang, H., Gao, C., Dantona, C., Hull, B., & Sun, J. (2024). DRG-LLaMA: Tuning LLaMA model to predict diagnosis-related group for hospitalized patients. *Npj Digital Medicine*, 7(1), 16. https://www.nature.com/articles/s41746-023-00989-3

Wu, Y., Li, Y., Baskys, A., Chok, J., Hoffman, J., & Roosan, D. (2024). Health disparity in digital health technology design. *Health and Technology*. https://doi.org/10.1007/s12553-024-00814-1

Yang, R., Tan, T. F., Lu, W., Thirunavukarasu, A. J., Ting, D. S. W., & Liu, N. (2023). Large language models in health care: Development, applications, and challenges. *Health Care Science*, 2(4), Article 4. https://doi.org/10.1002/hcs2.61